\documentclass[twoside,twocolumn,9pt]{article}
\usepackage{extsizes}
\usepackage[super,sort&compress,comma]{natbib} 
\usepackage[version=3]{mhchem}
\usepackage[left=1.5cm, right=1.5cm, top=1.785cm, bottom=2.0cm]{geometry}
\usepackage{balance}
\usepackage{mathptmx}
\usepackage{sectsty}
\usepackage{graphicx} 
\usepackage{lastpage}
\usepackage[format=plain,justification=justified,singlelinecheck=false,font={stretch=1.125,small,sf},labelfont=bf,labelsep=space]{caption}
\usepackage{float}
\usepackage{fancyhdr}
\usepackage{fnpos}
\usepackage[english]{babel}
\addto{\captionsenglish}{%
  
}
\usepackage{array}
\usepackage{droidsans}
\usepackage{charter}
\usepackage[T1]{fontenc}
\usepackage[usenames,dvipsnames]{xcolor}
\usepackage{setspace}
\usepackage[compact]{titlesec}
\usepackage{hyperref}

\usepackage{epstopdf}

\definecolor{cream}{RGB}{222,217,201}

\begin{document}

\pagestyle{fancy}
\thispagestyle{plain}
\fancypagestyle{plain}{
\renewcommand{\headrulewidth}{0pt}
}

\makeFNbottom
\makeatletter
\renewcommand\LARGE{\@setfontsize\LARGE{15pt}{17}}
\renewcommand\Large{\@setfontsize\Large{12pt}{14}}
\renewcommand\large{\@setfontsize\large{10pt}{12}}
\renewcommand\footnotesize{\@setfontsize\footnotesize{7pt}{10}}
\makeatother

\renewcommand{\thefootnote}{\fnsymbol{footnote}}
\renewcommand\footnoterule{\vspace*{1pt}%
\color{cream}\hrule width 3.5in height 0.4pt \color{black}\vspace*{5pt}} 
\setcounter{secnumdepth}{5}

\makeatletter 
\renewcommand\@biblabel[1]{#1}            
\renewcommand\@makefntext[1]%
{\noindent\makebox[0pt][r]{\@thefnmark\,}#1}
\makeatother 
\renewcommand{\figurename}{\small{Fig.}~}
\sectionfont{\sffamily\Large}
\subsectionfont{\normalsize}
\subsubsectionfont{\bf}
\setstretch{1.125} 
\setlength{\skip\footins}{0.8cm}
\setlength{\footnotesep}{0.25cm}
\setlength{\jot}{10pt}
\titlespacing*{\section}{0pt}{4pt}{4pt}
\titlespacing*{\subsection}{0pt}{15pt}{1pt}

\fancyfoot{}
\fancyfoot[LO,RE]{\vspace{-7.1pt}\includegraphics[height=9pt]{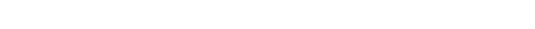}}
\fancyfoot[CO]{\vspace{-7.1pt}\hspace{13.2cm}\includegraphics{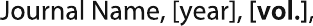}}
\fancyfoot[CE]{\vspace{-7.2pt}\hspace{-14.2cm}\includegraphics{RF}}
\fancyfoot[RO]{\footnotesize{\sffamily{1--\pageref{LastPage} ~\textbar  \hspace{2pt}\thepage}}}
\fancyfoot[LE]{\footnotesize{\sffamily{\thepage~\textbar\hspace{3.45cm} 1--\pageref{LastPage}}}}
\fancyhead{}
\renewcommand{\headrulewidth}{0pt} 
\renewcommand{\footrulewidth}{0pt}
\setlength{\arrayrulewidth}{1pt}
\setlength{\columnsep}{6.5mm}
\setlength\bibsep{1pt}

\makeatletter 
\newlength{\figrulesep} 
\setlength{\figrulesep}{0.5\textfloatsep} 

\newcommand{\topfigrule}{\vspace*{-1pt}%
\noindent{\color{cream}\rule[-\figrulesep]{\columnwidth}{1.5pt}} }

\newcommand{\botfigrule}{\vspace*{-2pt}%
\noindent{\color{cream}\rule[\figrulesep]{\columnwidth}{1.5pt}} }

\newcommand{\dblfigrule}{\vspace*{-1pt}%
\noindent{\color{cream}\rule[-\figrulesep]{\textwidth}{1.5pt}} }

\makeatother

\twocolumn[
  \begin{@twocolumnfalse}
{\includegraphics[height=30pt]{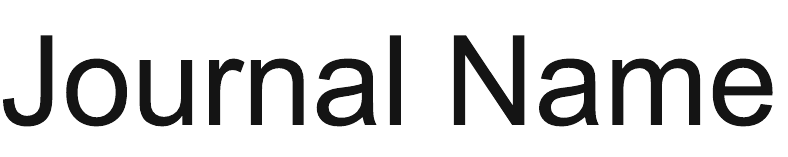}\hfill\raisebox{0pt}[0pt][0pt]{\includegraphics[height=55pt]{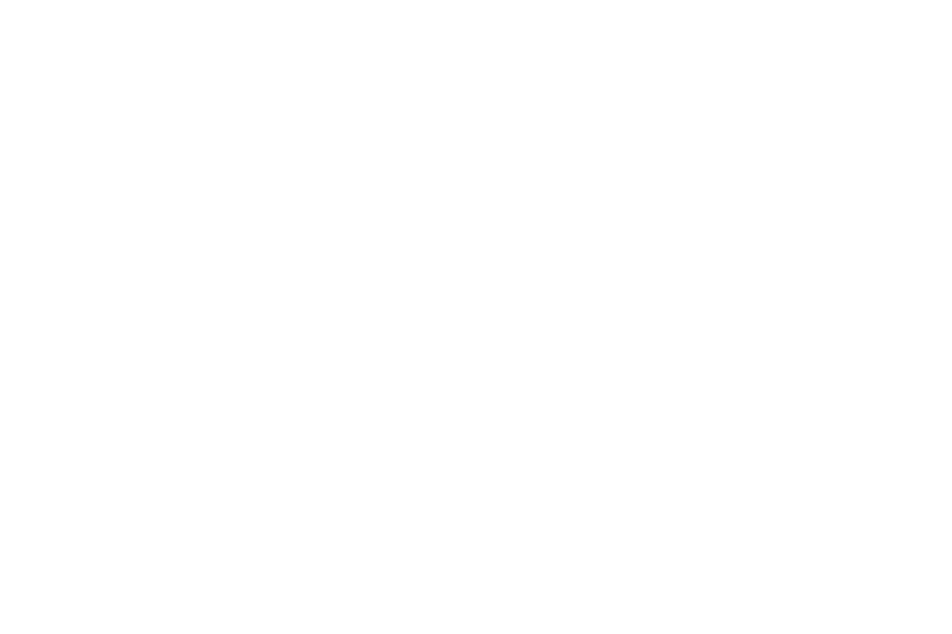}}\\[1ex]
\includegraphics[width=18.5cm]{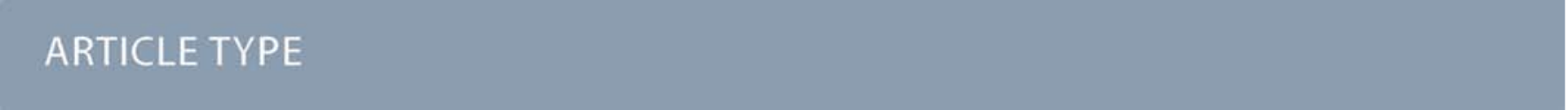}}\par
\vspace{1em}
\sffamily
\begin{tabular}{m{4.5cm} p{13.5cm} }

\includegraphics{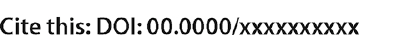} & \noindent\LARGE{\textbf{Electromechanics of the liquid water vapour interface}} \\
\vspace{0.3cm} & \vspace{0.3cm} \\

 & \noindent\large{Chao Zhang\textit{$^{a}$} and Michiel Sprik$^{\ast}$\textit{$^{b}$}} \\

\includegraphics{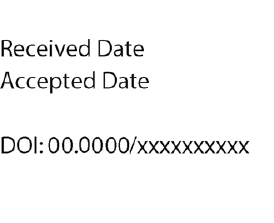} & \noindent\normalsize{Two collective properties distinguishing the thin liquid water vapour interface from the bulk liquid are the anisotropy of the pressure tensor giving rise to surface tension and the orientational alignment of the molecules leading to a finite dipolar surface potential. Both properties can be regarded as capillary phenomena and are likely to be coupled. We have investigated this coupling by determining the response of the tangential component of the surface tension to the application of an electric field normal to the surface using finite field molecular dynamics simulations. We find an upside down parabola with a maximum shifted away from zero field. Comparing the molecular dynamics results to an elementary electromechanical continuum model we relate the zero field derivative of the tangential part of the surface tension to the electrostatic potential generated by the spontaneous dipole alignment. The calculations show that these quantities have similar values but are not and in fact need not be identical. The electromechanical model also allows us to convert the absolute  curvature of the quadratic field dependence to an effective dielectric constant of the water interface which is found to be much lower compared to the bulk value as expected.} \\

\end{tabular}

 \end{@twocolumnfalse} \vspace{0.6cm}

]

\renewcommand*\rmdefault{bch}\normalfont\upshape
\rmfamily
\section*{}
\vspace{-1cm}


\footnotetext{\textit{$^{a}$~Department of Chemistry-\AA{}ngstr\"{o}m Laboratory, Uppsala University, L\"{a}gerhyddsv\"{a}gen 1, BOX 538, 75121, Uppsala, Sweden }}
\footnotetext{\textit{$^{b}$~Department of Chemistry, University of Cambridge, Cambridge CB2 1EW, United Kingdom. E-mail: ms284@cam.ac.uk}}




\section{Introduction}
\label{sec:intro}
As is obvious in view of its utmost importance, the liquid water vapour interface has been and continues to be a prime topic in molecular simulation and modelling. This started soon after it had been established that robust and relatively simple interaction site models (ISM's), such as SPC/E~\cite{Berendsen:1987uu} and TIP4P~\cite{Jorgensen1983}, are capable of giving a sufficiently accurate description of the structure and thermodynamics of bulk liquid water. Already in the early studies it was found that the dipoles of the water molecules in a narrow interfacial strip are aligned to give a net surface dipole\cite{Kataoka1988,Pratt1988}. The orientation is pointing inward from the vapour to the liquid. The alignment creates a dipole surface potential which should be distinguished from the overall interface electrostatic potential. While the dipole potential is physical, the precise value of the step in the overall Poisson potential is determined by nonphysical features of an ISM\cite{Pratt1992}. This picture was confirmed and refined in subsequent simulation studies\cite{Tildesley1997,Ichiye2015} although the value of the surface dipole potential $\chi^M$ showed some variation depending on the model\cite{Tildesley1997}. Our value for SPC/E, the model used in the present application, is 320 meV, which is in good agreement with the estimates of other authors for the same model.          

While well defined for an ISM, it is not clear how to extract the dipole surface potential from experimental measurement. This rather disconcerting complication is particularly relevant for electrochemistry (we will defer a discussion of this difficult issue to section \ref{sec:disc}). On the other hand, there is a surface property that is accessible to experiment, namely the surface tension. After some technical issues concerning long-range electrostatic interactions had been settled, the experimental surface tension of the liquid water vapour interface (720 bar*nm under ambient conditions) could be reproduced by ISM based molecular dynamics (MD) simulation within a 10 percent error\cite{Tildesley1995,Dhir2006,Grest2006,Vega2007,Jackson2015,Jedlovszky2016,Nikzad2017} (for a review see Ref.~\citenum{Tildesley2016}). The value we obtained using SPC/E is 610 bar*nm (see Fig.~\ref{fgr:system} in Section ~\ref{sec:results}).     

The aim of the present contribution is to investigate whether there is a quantitative electromechanical relation between surface potential and surface tension of the liquid water vapour interface. Such a relation is suggested by a comparison of the narrow width of the anisotropy in the pressure tensor and the alignment of molecular dipoles. According to mechanical theory, surface tension is generated by a difference between  the normal and lateral component of the pressure tensor\cite{Rowlinson1982}. The expression for the surface tension $\gamma$ for a planar interface is
\begin{equation}
  \gamma = \int_{z_{v}}^{z_{l}}dz \left(p_N - p_T(z)\right)
  \label{eq:gamplan}
\end{equation}
where $p_N$ is the pressure along the $z$ direction chosen normal to the surface and $p_T$ is the tangential component of the pressure tensor. $z_l > z_v$ are coordinates in the bulk region of the liquid respectively vapour, where $p_N = p_T$. Mechanical equilibrium requires that $p_N(z)=p_N$ is equal to the bulk value independent of $z$.  The interface region is under tension with $p_T(z) < 0 $. For the liquid water vapour interface $p_T$ reaches a minimum value of as low as $-2$ kbar under ambient conditions ($p_N = 1$ bar)\cite{Jedlovszky2016}. However the interval over which $p_T \ne p_N$ is only about 0.5 nm wide (see Fig.~3 of Ref.~\citenum{Jedlovszky2016}). This is very close to the interval over which the water molecules in the interfacial layer  show a net orientation (see e.g.~Fig.~3 of Ref.~\citenum{Ichiye2015}). This observation suggests that the surface potential, similar to the surface tension, is a capillary effect. Viewing the surface potential from a mechanical perspective might therefore help understanding the dipole alignment at the interface.

The mechanism controlling the orientation of interfacial water dipoles is indeed still somewhat of a mystery. Aligning dipoles in a thin sheet costs energy. Clearly another mechanism than dipole-dipole coupling must be responsible. Ideally a chemist would like to explain this interaction in terms of the intricate pattern of hydrogen bonding. While this may be the ultimate answer, the question has not been resolved yet to every bodies satisfaction. Here liquid state density functional theory (DFT) calculations have provided useful input\cite{Gray1992,Gray1994,Dietrich1992,Dietrich1993,Warschavsky2002,Warschavsky2003}. Rather than ISM Hamiltonian's, these calculations are based on point dipole+quadrupole interaction models. The picture emerging from this work is that the electric field induced by the ordering of the quadrupoles aligns the dipoles. This is consistent with the general view that quadrupoles play a dominant role in multipole modelling of water. The question remains of course what is ordering the quadrupoles.

As we argued above continuum thermomechanics may be able to add a different and complementary perspective. The centrepiece in the continuum theory of capillarity is the VanderWaals square gradient free energy model. The key insight of VanderWaals was that finite surface tension is due to the steep gradient in the density at the interface\cite{Rowlinson1982}(see also Ref.~\citenum{Gibbs1976}). The VanderWaals theory can be regarded as a form of Landau-Ginzburg theory for non-uniform liquids. Gradients are also central in the Landau-Ginzburg theory of non-uniform solids, but these gradients are spatial derivatives  of strain. This makes the formalism more complicated but the reason that the continuum mechanics of solids should be of interest for our problem is that the coupling to electromagnetic fields has been investigated in depth in this field\cite{Ogden2005,Ogden2017,Suo2008,Suo2010}(see also Ref.~\citenum{Maugin1988}). The theory of non-uniform elasticity has recently received a boost when it was realised that strain gradients break inversion symmetry in centro-symmetry materials which now are also susceptible to polarization in response to strain just as non-centro symmetric piezo electric systems are. This is the phenomenon of flexoelectricity which has been observed in hard ceramic crystals \cite{Tagantsev2013,Sharma2008, Shen2010} as well as soft molecular and even biological systems\cite{Meyer1969,Sharma2014}. The implication for us is that the spontaneous polarization at the liquid water vapour interface could possibly also be viewed as a flexoelectric effect induced by the gradient in density.

The work reported here is a preliminary exploration of electromechanical coupling at the liquid water vapour interface.  We study the response of the surface tension to the application of an electric field normal to the interface. The motivation was the hypothesis of a link to gradient continuum electroelasticity theory, as outlined above. However, we are not yet ready for this  and we must leave this as speculation at this stage (for further discussion see the outlook in section \ref{sec:disc}). Still continuum electromechanics plays an important role in this paper. The finite field molecular dynamics simulations are analyzed in terms of the simplest of electroelastic models, a uniform dielectric membrane (no gradients). This model is taken from the literature on dielectric elastomers\cite{Pelrin2000,Kofod2003,Ogden2017}. Electromechanical coupling is accounted for by uniform Maxwell stress only. Our key result is a simple relation between the interface potential and the zero field derivative of the (lateral) surface tension. This is the main distinction between the present contribution and earlier molecular dynamics studies of the electromechanical response of the liquid vapour interface of water\cite{Grest2006,Nikzad2017}

The organization of the paper is as follows. The electromechanical model used for the interpretation of our results is outlined in section \ref{sec:elemech} with only minimal justification. Section \ref{sec:method} is the technical section describing the finite field molecular dynamics  scheme and the computation of the pressure tensor in combination with Ewald summation for electrostatic interactions. The specification of the model system can also be found in section \ref{sec:method}. The molecular dynamics results are presented in section \ref{sec:results}. As will become clear this exploratory study leaves a large number of loose ends, both regarding theory and computational method imposing restrictions on the interpretation of the results. These limitations are summarized in section \ref{sec:disc} together with an outlook of how these issues might be addressed.

\section{Continuum electromechanics} \label{sec:elemech}

\subsection{Choice of dielectric Maxwell stress tensors}
\label{sec:Mstress}
The Maxwell stress tensor $\mathbf{\sigma}^{\mathrm{M}}$ is
a second rank tensor quantifying the stress generated by long-range electrostatic forces. Adding the stress tensor $\mathbf{\sigma}$ due to all other supposedly short-range forces gives the total stress tensor
\begin{equation}
\mathbf{\sigma}^{\mathrm{t}}= \mathbf{\sigma} +\mathbf{\sigma}^{\mathrm{M}}
\label{eq:tstress}
\end{equation}
controlling mechanical equilibrium. $\mathbf{\sigma}^{\mathrm{t}}$ must therefore be symmetric (2nd Cauchy law).  
 
The difficulty, discussed at length in the literature, is that there is no unique way of partitioning the stress in electrostatic and short-range contributions. In the literature on non-uniform polar fluids the Maxwell stress tensor is often given in the form referred to as the Kelvin stress tensor $\mathbf{\sigma}^{\mathrm{K}}$
\begin{equation}
  \mathbf{\sigma}^{\mathrm{K}} = \frac{1}{4 \pi} \left(
  \mathbf{D} \otimes \mathbf{E}
  - \frac{1}{2} \left(\mathbf{E} \cdot \mathbf{E} \right) \mathbf{I} \right) 
  \label{eq:MKelv}
\end{equation}
where $\mathbf{I}$ is the unit tensor. $\mathbf{D}$ is the electric displacement field and $\mathbf{E}$ is the Maxwell electric field. The difference is the polarization
\begin{equation}
  \mathbf{D} = \mathbf{E} + 4 \pi \mathbf{P}
\label{eq:DEP}
\end{equation}
Note that, following Landau and Lifshitz\cite{Landau1984}, we use the Gaussian system of electrical units. Note also that the Kelvin stress tensor Eq.~\ref{eq:MKelv} is not symmetric and already for this reason cannot be identified with the total stress tensor (Eq.~\ref{eq:tstress}).  

Arguments in the literature on dielectric elastic solids, on the other hand, are usually based on a Maxwell stress tensor explicitly dependent on the relative dielectric constant $\epsilon$ of the material 
\begin{equation}
  \mathbf{\sigma}^{\mathrm{E}} = \frac{\epsilon}{ 4 \pi}
  \bigl( \mathbf{E} \otimes \mathbf{E} - 
  \frac{1}{2} \left(\mathbf{E} \cdot \mathbf{E} \right) \mathbf{I} \bigr)
\label{eq:MHelmE}
\end{equation}
with
\begin{equation}
  \mathbf{D} = \epsilon(\mathbf{r})  \mathbf{E}
  \label{eq:DepsE}
\end{equation}
We have allowed for a local electric constant $\epsilon=\epsilon(\mathbf{r})$
depending on position $\mathbf{r}$. The spatial variation is due to an implicit dependence on $\mathbf{r}$ because dielectric constants $\epsilon=\epsilon(\rho)$ in linear dielectrics generally change with number density $\rho(\mathbf{r})$. The tensor Eq.~\ref{eq:MHelmE} is a simplified form of the Korteweg-Helmholtz stress tensor (see below).

While of similar appearance $\mathbf{\sigma}^{\mathrm{K}}$ and $\mathbf{\sigma}^{\mathrm{E}}$ are not equivalent. The distinction becomes more manifest when comparing the corresponding force densities. $\mathbf{\sigma}^{\mathrm{K}}$ generates the Kelvin force density acting on polarization subject to a nonuniform electric field
\begin{equation}
  \mathbf{f}^{\, \mathrm{K}}=\nabla \cdot \mathbf{\sigma}^{\mathrm{K}} = \mathbf{P} \cdot \nabla \mathbf{E}
\label{eq:fdK}
\end{equation}
Viewed from a microscopic perspective $\mathbf{f}^{\, \mathrm{K}}$ is the  force per volume on a system of point dipoles with dipole density $\mathbf{P}$ (See e.~g. Ref.~\citenum{Melcher1981}). The divergence of $\mathbf{\sigma}^{\mathrm{E}}$ of Eq.~\ref{eq:MHelmE} yields
\begin{equation}
  \mathbf{f}^{\, \mathrm{E}} = \nabla \cdot \mathbf{\sigma}^{\mathrm{E}} = 
  - \frac{\mathbf{E}^2}{8 \pi} \nabla \epsilon 
\label{eq:fdE}
\end{equation}
which looks rather different from Eq.~\ref{eq:fdK}. What is the difference?
First of all the Kelvin force density Eq.~\ref{eq:fdK} is more general. To derive it we need the polarization equation Eq~\ref{eq:DEP} valid in every system and the Maxwell equations for the electric field.
\begin{equation}
  \nabla \wedge \mathbf{E}   =  0
\label{eq:curlE}
\end{equation}
and displacement (in absence of free and external charge) 
\begin{equation}
  \nabla \cdot \mathbf{D}  =  0
\label{eq:divD}
\end{equation}
The constitutive relation Eq.~\ref{eq:DepsE}  is not required. On the other hand to obtain the force density Eq.~\ref{eq:fdE} we must impose  Eq.~\ref{eq:DepsE} in addition to the Maxwell field equations Eqs.~\ref{eq:curlE} and \ref{eq:divD}.

\subsection{Which Maxwell stress tensor to use}
\label{sec:Mchoice}
We will now argue that the stress tensor of Eq.~\ref{eq:MHelmE}  is the more suitable for a first qualitative analysis of our MD results. The Kelvin force density Eq.~\ref{eq:fdK} is of course also defined for a linear dielectric satisfying the constitutive relation eq.~\ref{eq:DepsE} and it should be possible to directly compare to the force density of Eq.~\ref{eq:fdE}. Indeed, for a finite body for which surface contributions can be made to vanish it can be shown\cite{Melcher1981,Zahn2006} that $\mathbf{f}^{\, \mathrm{K}}$ and $\mathbf{f}^{\, \mathrm{E}}$ differ by the gradient of a scalar field
\begin{equation}
\mathbf{f}^{\, \mathrm{E}} = \mathbf{f}^{\, \mathrm{K}} - \nabla p^{\, \mathrm{E}}
\label{eq:DfKfE}
\end{equation}
$p^{\, \mathrm{E}}$  can be interpreted as an electrostatic pressure or the stored polarization energy
\begin{equation}
 p^{\, \mathrm{E}} =  \left( \epsilon - 1 \right)  
 \frac{\mathbf{E}^2}{8 \pi} = \frac{\mathbf{P}^2}{2 \chi}
 \label{eq:pE}
\end{equation}
where $\chi = (\epsilon-1)/(4 \pi)$ is the susceptibility. The combination of
Eqs.~\ref{eq:DfKfE} and \ref{eq:pE} suggests that while $\mathbf{f}^{\, \mathrm{K}}$ and    
$\mathbf{f}^{\, \mathrm{E}}$ both account for the phoretic force exerted by a non-uniform electric field, $\mathbf{f}^{\, \mathrm{E}}$ in addition also includes a thermodynamic contribution related to an inhomogeneous dielectric free energy density.  

The more inclusive thermodynamic status of $\sigma^{\mathrm{E}}$ is confirmed by a proper electromechanical derivation of the stress tensor in a linear dielectric. This derivation however extends $\sigma^{\mathrm{E}}$ of Eq.~\ref{eq:MHelmE} with a further term due to electrostriction. 
\begin{equation}
  \mathbf{\sigma}^{\mathrm{H}} =
  \epsilon \frac{ \mathbf{E} \otimes \mathbf{E}}{ 4 \pi}  -
   \frac{\mathbf{E}^2}{ 8 \pi} \left(
   \epsilon - \rho \frac{\partial \epsilon}{ \partial \rho} \right)  \mathbf{I}
\label{eq:MHelmK}
\end{equation}
with the corresponding extended force density 
\begin{equation}
  \mathbf{f}^{\, \mathrm{H}} =  - \frac{\mathbf{E}^2}{8 \pi} \nabla \epsilon + \frac{1}{8 \pi} 
 \nabla \left( \rho \frac{\partial \epsilon }{\partial \rho} \frac{\mathbf{E}^2}{2} \right)
 \label{eq:fdHK}
\end{equation}
$\mathbf{f}^{\, \mathrm{H}}$ of Eq.~\ref{eq:fdHK} is known as the Korteweg-Helmholtz force density\cite{Landau1984,Melcher1981,Zahn2006}. The generating stress tensor Eq.~\ref{eq:MHelmK} was obtained by Landau and Lifshitz using work arguments\cite{Landau1984}. We should also mention the profound contribution of McMeeking and Landis who approach the problem of the definition of the Maxwell stress using a virtual work rate principle from (quasistatic) non-equilibrium continuum thermodynamics\cite{Landis2005}.  Alternatively the same Korteweg-Helmholtz tensor can be obtained applying a continuum mechanics Lagrangian scheme\cite{Suo2008,Suo2010}. This is perhaps the more illuminating for our purpose.

The central degree of freedom in the  continuum theory of elasticity is the deformation of a solid relative to a reference or material frame. The stress tensor is computed as a   deformation derivative of a mechanical energy density designed to reproduce the relevant constitutive stress-strain relations. The basics of linear elasticity were already established in 19 the century. In a subsequent development the theory was extended to non-linear (hyper) elasticity which then allowed for coupling to electric field (as indicated by the quadratic dependence on electric fields of the Maxwell stress tensor the electromechanics is intrinsically non-linear)\cite{Ogden2005,Suo2008,Landis2005}. A formal presentation of continuum electromechanics relies heavily on tensor algebra which will be rather out of place in the present paper (for a recent review and references see e.g~Ref.\citenum{Ogden2017}).    

A liquid has no memory. Free energy densities are dependent on the current configuration only.  However, as pointed out by Suo et al\cite{Suo2008} a reference frame does not have to be physical and it is still possible to obtain an expression for a electromechanical stress tensor of a liquid by a derivative of deformation relative to an arbitrary reference frame. In Ref.~\citenum{Suo2008} this was carried out for an ideal dielectric fluid given by the free energy density
\begin{equation}
  f(\rho, \mathbf{D}) = f_0(\rho) + \frac{\mathbf{D}^2}{8 \pi \epsilon}
  \label{eq:fideal}
\end{equation}
where $\rho$ is the density of the fluid and $\mathbf{D}$ is again the dielectric displacement. $f_0$ is the free energy density function at zero field. Transforming the free energy density Eq.~\ref{eq:fideal} to an arbitrary reference frame introduces the tensor describing the deformation as an additional independent degree of freedom. Evaluating the derivative and transforming back to the current gives a total stress tensor of the form
\begin{equation}
\mathbf{\sigma}^{\mathrm{t}}= -p_0 \mathbf{I} +\mathbf{\sigma}^{\mathrm{H}}
\label{eq:sideal}
\end{equation}
where $p_0$  the pressure derived from $f_0$ and $\mathbf{\sigma}^{\mathrm{H}}$ the dielectric Korteweg-Helmholtz stress of Eq.~\ref{eq:MHelmK}.

This somewhat counter intuitive derivation based on mathematical manipulation in a fictitious reference system brings out a unique feature of electromechanical coupling. The only apparent coupling between density and electric field in Eq.~\ref{eq:fideal} is a possible density dependence of the dielectric constant (electrostriction). However, as the experiments on elastomers showed, also a uniform incompressible dielectric elastic slab responds to application of electric fields by stretching. It does this even though according to Eq.~\ref{eq:fdHK} the (bulk) force density should vanish in this limit. As clearly explained by Suo et al\cite{Suo2008,Suo2010} using the example of a parallel plate capacitor this is a manifestation of geometric coupling controlled by the residual Maxwell stress tensor $\mathbf{\sigma}^{\mathrm{E}}$ of Eq.~\ref{eq:MHelmE}.       

\subsection{Dielectric membrane model and analysis of MD results}
\label{sec:dielbrane}
The key requirement for a model for the electromechanical behaviour of the water vapour interface is that it should account for the quadratic response to an applied electric field as observed in our simulation (the upside down parabola). Electrostriction is such a second order effect (see eq.~\ref{eq:MHelmK}). But, as argued in section \ref{sec:Mchoice} ignoring electrostriction still leaves the geometric coupling which also leads to quadratic response (Eq.~\ref{eq:MHelmE}).  This is the minimal model we will adopt. The water vapour interface is treated as an incompressible membrane. We are aware that the gradient in the interfacial density is the essence of capillarity\cite{Rowlinson1982} and leaving this out is a most serious simplification.  This and further obvious limitations of our primitive model will discussed in section \ref{sec:disc}.  

The external field $\mathbf{E}_0$ is applied normal to the flat interface. The displacement field $\mathbf{D}$ is uniform and equal to $\mathbf{E}_0$ everywhere even if the electric field $\mathbf{E}$ is not uniform. Substituting the constitutive relation Eq.~\ref{eq:DepsE} in Eq.~\ref{eq:MHelmE} we can write \begin{equation}
  \mathbf{\sigma}^{\mathrm{E}} = \frac{1}{ 4 \pi}
  \bigl( \mathbf{D} \otimes \mathbf{E} - 
  \frac{1}{2} \left(\mathbf{D} \cdot \mathbf{E} \right) \mathbf{I} \bigr)
\label{eq:MHelmD}
\end{equation}

\begin{gather}
  \sigma^{\mathrm{E}}_N = \sigma^{\mathrm{E}}_{zz} = \frac{D_z E_z}{8 \pi}
  \label{eq:sigmaM}\\
  \sigma^{\mathrm{E}}_T = \sigma^{\mathrm{E}}_{xx} = \sigma^{\mathrm{E}}_{yy} =
  - \frac{D_z E_z}{8 \pi}
  \label{eq:sigmaT}
\end{gather}
Omitting the $z$ coordinate subscript ($D= D_z, E=E_z$ and we have for the component of the stress tensor parallel to the surface
\begin{equation}
\sigma_T = -p_T =   -\frac{DE}{8 \pi \epsilon }
\label{sigDE}
\end{equation}
In our primitive membrane model the electric fields are assumed to be uniform over the width of the interface layer. Integrating over the interface layer amounts to multiplication by the width $l$ of the interface. Setting $E=D/\epsilon$ we find for the lateral component to the surface tension
\begin{equation}
  \gamma_T = \gamma^{0}_{T} -\frac{l D^2}{8 \pi \epsilon }
\label{eq:gamiel}
\end{equation}
$\gamma^{0}_{T}$ stands for the contribution of the short-range forces (see Eq.~\ref{eq:sideal}. If the membrane is effectively incompressible $\gamma^{0}_{T}$ is not affected by the application of the electric field neither is the dielectric constant $\epsilon$. Under these rather assumptions the $D$ dependence of Eq.~\ref{eq:gamiel} is indeed an upside down parabola.

\begin{equation}
  E= \frac{D}{\epsilon} - 4 \pi  P_0
\label{eq:EDP0}  
 \end{equation} 

\begin{equation}
  \frac{ d P_0}{d D} = 0
\label{eq:dP0dD} 
\end{equation}

The quantity $\chi^0 = 4 \pi P_0 l$ can be interpreted as the potential difference across the interface generated by the spontaneous polarization. 
\begin{equation}
  4 \pi \left. \frac{d \gamma_{\,T}}{d D} \right|_{D=0} = \chi^0
\label{eq:chi0}
\end{equation}

Since we did not allow for any $D$ dependence of the spontaneous polarization the second derivative of the $ \gamma_{\,T}(D)$ parabola is simply determined by the effective dielectric constant of the later and its width
\begin{equation}
\frac{d^2 \gamma_{\,T}}{d D^2} = - \frac{l}{4 \pi \epsilon}
\label{eq:epsilon}
\end{equation}
$d^2 \gamma_{\,T}/d D^2$ is a quantity with the dimensions of length. Thus introducing the characteristic length
\begin{equation}
\lambda =  4 \pi \left| \frac{d^2 \gamma_{\,T}}{d D^2} \right|
\end{equation}
we can write
\begin{equation}
  \epsilon = \frac{l}{\lambda}
 \label{eq:leps}
\end{equation}
Recall that $\epsilon $ in Eq.~\ref{eq:leps} is an effective dielectric constant of the interface as defined in the membrane model. Its value can therefore be expected to be significantly smaller than the value of the dielectric constant of approximately 80 in bulk water.

Eq.~\ref{eq:chi0} and \ref{eq:leps} are the two central relations suggested by the membrane model for analysis of the MD results.  

\section{Molecular dynamics}
\label{sec:method}
\subsection{Finite field Hamiltonian and model system}
\label{sec:vdbslab}
Computing the finite field derivative of the surface tension was carried out by finite field MD simulations~\cite{Zhang2016,Zhang2019} of the liquid water vapour interface at ambient conditions with the hybrid constant electric displacement $D$ Hamiltonian. ``Hybrid'' means the $D$ field and the corresponding electric boundary condition are only applied in the direction perpendicular to the surface. The extended Hamiltonian is written as
\begin{equation}
\label{eq:uvdb}
H_D\left(v ,D \right) = H_{\mathrm{PBC}}(v)+ \frac{\Omega}{8 \pi} 
  \left(D - 4 \pi P(v) \right)^2
\end{equation}
where $P$ is the itinerant polarization in the direction of $D$ (See Secs. IV B and IV C in Ref. ~\citenum{edl2016} for the elaboration). $\Omega = L_x L_y L_z$ is the supercell volume and $v= (\mathbf{r}^N ,\mathbf{p}^N)$ stands for the collective momenta and position coordinates of the $N$ particles in the system. We will use a system geometry in which the $z$ axis is perpendicular to the water slab (Fig.~\ref{fgr:system}b). In this geometry $P$ in Eq.~\ref{eq:uvdb} is the polarization of the full supercell and is therefore equal to $M/\Omega$ where $M$ is the instantaneous dipole moment of the water slab (see again Fig.~\ref{fgr:system}b). Writing the MD Hamiltonian Eq.~\ref{eq:uvdb} in terms of the dipole moment $M$ we have \begin{equation}
\label{eq:uvdbM}
H_D\left(v ,D \right) = H_{\mathrm{PBC}}(v)+  \frac{\Omega }{8 \pi}D^2 -  DM(v) +  \frac{2 \pi }{\Omega} M(v)^2
 \end{equation}
 Setting $D=0$ in Eq.~\ref{eq:uvdbM} we recognize the Yeh-Berkowitz dipole screened slab Ewald Hamiltonian\cite{Yeh1999}. Finite $D$ adds the expected coupling between the dipole moment and the applied field and (the $-DM$ term) plus a (constant) vacuum energy of the applied field. We will return to Eq.~\ref{eq:uvdbM} when analyzing finite system size effects.   

The simulation system consists of 706 water molecules and a vacuum slab in a fixed rectangular box of 2.77 nm$\times$2.77 nm $\times L_z$ ($L_z$=5.55 nm, 8.33 nm and 11.11 nm). The interactions are described by the SPC/E model potential~\cite{Berendsen:1987uu}. The molecules are kept rigid using the SHAKE algorithm~\cite{Ryckaert:1977gp}. The MD integration time step is 2~fs and the trajectory is 20~ns-long at each condition (box size and $D$ field strength). The Ewald summation is implemented using the Particle Mesh Ewald (PME) scheme\cite{Ewald}. Short-range cutoffs for the VanderWaals and Coulomb interaction in the direct space are 1.2 nm. The temperature is controlled by a Nos\'e-Hoover chain thermostat~\cite{martyna92} at 298K. 

\subsection{Pressure tensor and surface tension calculation}
\label{sec:pTcalc}
The local pressure tensor $p_{\alpha \beta}$ with $\alpha, \beta = x,y,z$ required for the computation of the surface tension according to Eq.~\ref{eq:gamplan} is one of the more enigmatic observables in molecular dynamics simulation with a long history of debate\cite{Tildesley2016}. When obtained from a momentum balance equation the expression for $p_{\alpha \beta} $ is not unique (see Ref.~\citenum{Tildesley2016} for references to the original literature). For systems with short-range pair interactions such as Lennard-Jones the expression derived by Irving and Kirkwood (IK) is considered the most suitable giving results consistent with experiment and statistical mechanical theory\cite{Rowlinson1982}.

Long-range electrostatic interactions when computed using Ewald summation cannot be resolved in pair interactions (The  electromechanical argument of section \ref{sec:elemech} is a strong indication that this difficulty is not specific to Ewald summation but a generic property of electrostatics in extended systems). The surface tension for water can therefore not be computed using the IK pressure tensor. The solution proposed by Sonne et al.~is to use an alternative expression for  the pressure tensor due to Harasima\cite{Peters2005} (see also Ref.~\citenum{Tildesley2016}). The Harasima procedure was adapted by Sega et al.~for Particle Mesh Ewald (PME)\cite{Sega2016}. This is the method applied in the present calculation. Alternatively, the surface tension of water can be computed using a rather different thermodynamic approach avoiding calculation of force virials\cite{Vega2007,Jackson2015}.  

The scheme for calculation of the pressure tensor under Ewald periodic boundary conditions according to Ref.~\citenum{Peters2005} has unfortunately a number of restrictions. Interfaces must be planar and the method only gives an estimate of the tangential component $p_T$. It is not possible to obtain the normal component $p_N$. Normally calculation of $p_N$ is not necessary for planar interfaces $p_N$ being equal to the bulk pressure under conditions of mechanical equilibrium. However force balance under finite field is more involved (see e.g.~Ref.~\citenum{Ogden2005}). For example the contribution of the electric field (the Maxwell stress tensor) does not vanish in vacuum. In fact it is not immediately clear whether Eq.~\ref{eq:gamplan} is even valid under finite field (we return to this question in section \ref{sec:disc}).   

Clearly computation of surface tension under finite field is more complicated. To make a start with the MD study of the electromechanical response of the liquid water vapour interface we have decided to leave these complications as unresolved and only compute the tangential pressure profile. The integral  
\begin{equation}
    \label{gamma_prof}
    \gamma_T(z) = -\int_{-L_z/2}^z ds~p_T(s)
\end{equation}
with $p_T(s) = (p_{xx}(s)+p_{yy}(s))/2$ can therefore not be treated as an estimation of the surface tension except for the $D=0$ case.  All simulations for computing the local pressure components $p_{xx}(s)$ and $p_{yy}(s)$ were carried out with the virial version of GROMACS 4 package~\cite{Hess2008} from Sega~\cite{Sega2016,Jedlovszky2016} (\href{https://github.com/Marcello-Sega/gromacs/tree/virial}{https://github.com/Marcello-Sega/gromacs/tree/virial}). Note that the $\gamma_T(z)$ computed this way only contains the stress due to $H_{\mathrm{PBC}}$ in Eqs.~\ref{eq:uvdb} and \ref{eq:uvdbM}. The stress due to the constant $D$ extension is not included.  

\section{Results}
\label{sec:results}

\begin{figure}[h]
\centering
  \includegraphics[width=1.0\columnwidth]{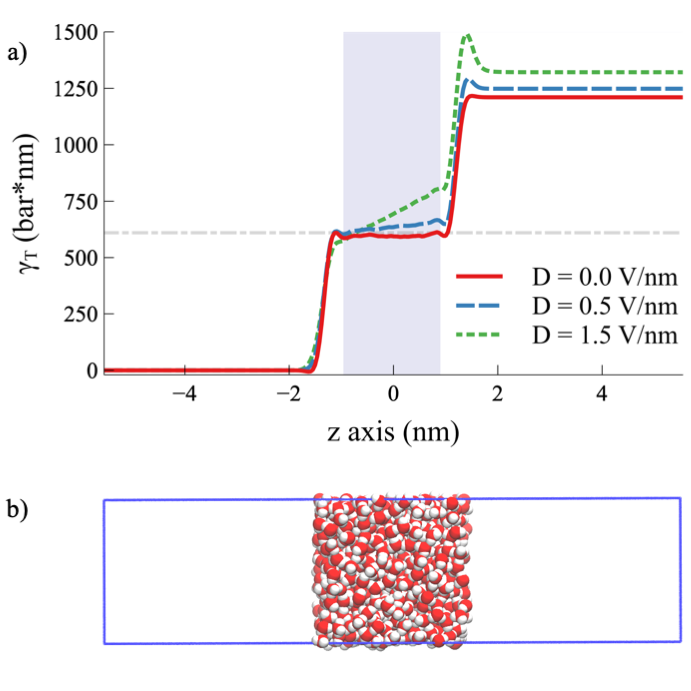}
  \caption{ a) Field-dependent profiles of the integral transverse component of surface tension $\gamma_T(z)$ as defined in Eq.~\ref{gamma_prof}. The dashed-dotted line is the reference value of the macroscopic $\gamma_T$ at $D=0.0$ V/nm amounting to 610 bar*nm for our model system. The shaded area indicates the bulk region of the water slab. b) A snapshot of the liquid water vapour model system used in our simulations. The total length of the MD box is $L_z = 11.11$ nm corresponding to the large box referred to in Fig.~\ref{fgr:derivative}. }
  \label{fgr:system}
\end{figure}

Fig.~\ref{fgr:system} shows the field-dependent profile of the transverse component of the surface tension calculated according to Eq.~\ref{gamma_prof}. The bulk value of $\gamma_T(z)$ is about 605 bar*nm which is quite close to the macroscopic $\gamma_T$ of 610 bar*nm (the dash-dotted line in Fig.~\ref{fgr:system}). Keeping in mind that $\gamma_T$ cannot be identified with the full surface tension at finite field as discussed in section \ref{sec:pTcalc}, the agreement at zero field provides confidence for the accuracy of the field-dependent $\gamma_T(z)$ profiles. 

\begin{figure}[h]
\centering
  \includegraphics[width=1.0\columnwidth]{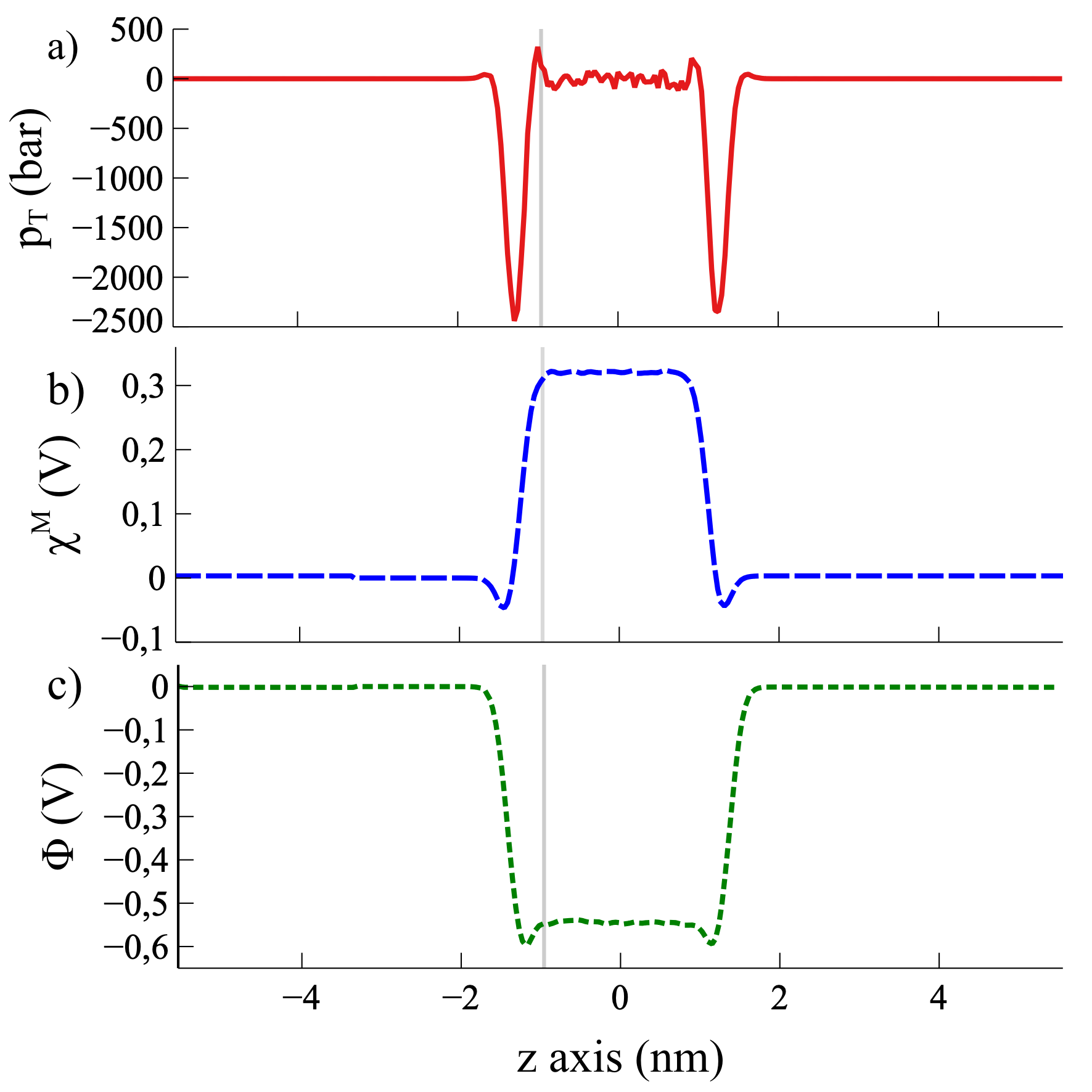}
  \caption{ The $D=0$ profile of a) the tangential component of the pressure tensor, b) the surface dipole potential $\chi^M$, c) the electrostatic potential $\Phi$. The grey vertical line indicates the dividing surface we chose for the estimation of the tangential component of the surface tension shown in Fig~\ref{fgr:derivative}.}
  \label{fgr:dividing}
\end{figure}

As shown in Fig.~\ref{fgr:system}a, the applied  field (equal to $D$ for our system) induces both surface and bulk response. This means one has to choose a dividing surface in order to separate the surface from the bulk. Our choice is 0.9 nm from the (average) location of the surface. This is the point where the tangential pressure  $p_T(z)$ starts to oscillate around zero at $D=0$ (see Fig.~\ref{fgr:dividing}a). This interval includes the topmost three layers of interfacial water molecules (Fig.~\ref{fgr:system}b) consistent with the detailed analysis  of Ref.~\citenum{Jedlovszky2016}. While this rule-of-thumb choice will of course affect the actual value of the field derivative (Eq.~\ref{eq:chi0}), it does not change the physical picture proposed in this work.

\begin{figure}[h]
\centering
  \includegraphics[width=1.0\columnwidth]{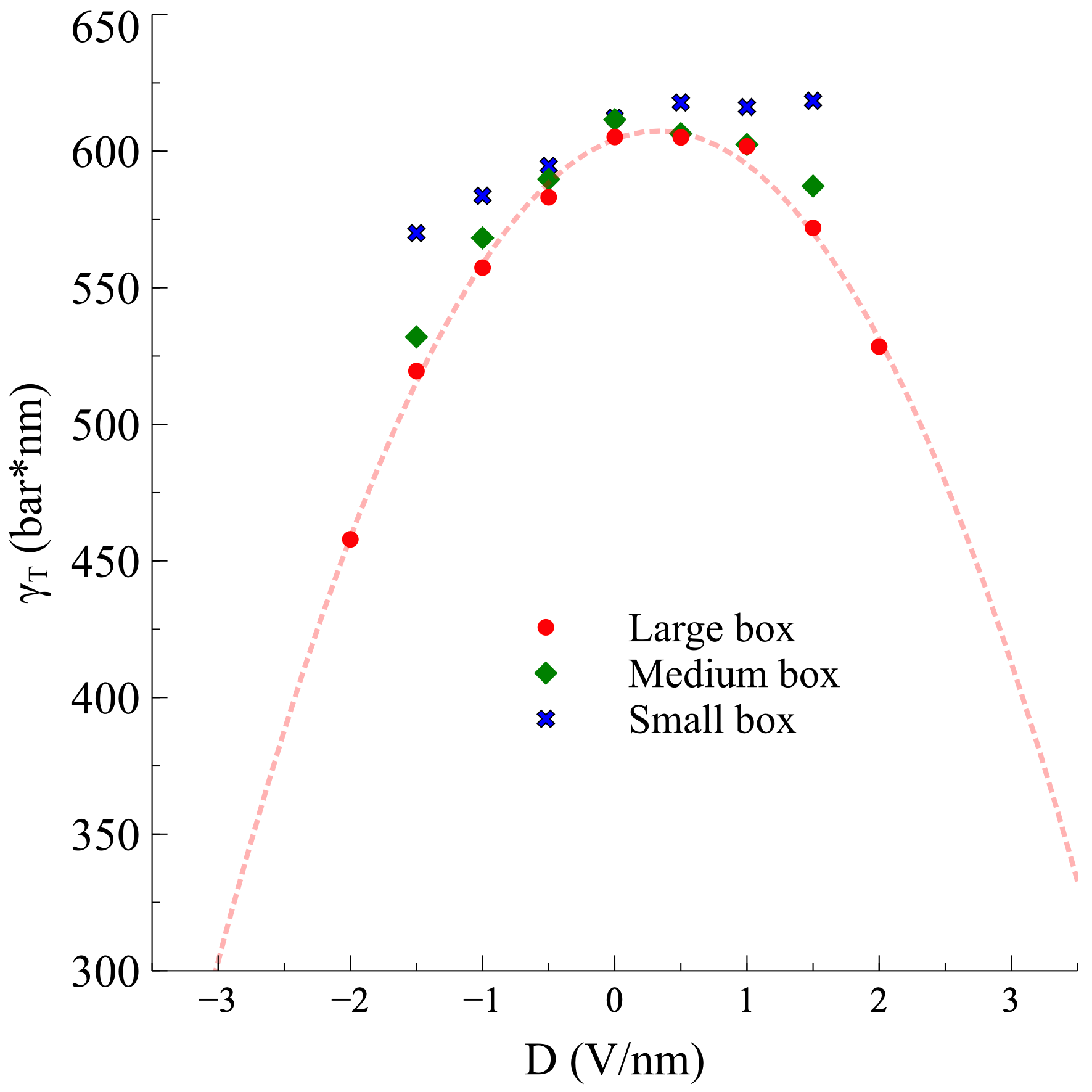}
  \caption{ Field-dependent tangential component  $\gamma_T$ of the surface tension  of the liquid water vapour interface. $L_z$=5.55 nm, 8.33 nm and 11.11 nm are the values for the length in the direction normal to the surface for a small, medium and large MD box respectively. A dividing surface is chosen to extract the value of $\gamma_T$ for the interface, which is about 0.9 nm from the onset of the tangential pressure $p_T(z)$ and shown by the vertical line in Fig.~\ref{fgr:dividing}.}
  \label{fgr:derivative}
\end{figure}

The resulting transverse component of surface tension $\gamma_T$ is given as a function of the field strength  in Fig.~\ref{fgr:derivative}. Three different values of the length $L_z$ of the MD box normal the surface were investigated. Clearly there is a dependence on system size, which will be addressed at the end of this section after discussion of the main results. Focusing on the large box curve (in red) in Fig~\ref{fgr:derivative} we notice two features: i) The field-dependence is essentially a concave parabola opening downward; ii) There exists a linear term which makes the parabola asymmetric with respect to $D=0$. This shape is consistent with the spontaneously polarized dielectric membrane model of section \ref{sec:dielbrane}. Hence fitting the simulation results to a shifted quadratic curve and converting the parameters to the zero field derivative of Eq.~\ref{eq:chi0} and effective dielectric constant of Eq.~\ref{eq:epsilon}, we find  $\chi^0 \approx $ 210 mV and $l/(4\pi\epsilon) \approx$ 0.025 nm respectively. 

The surface dipole potential $\chi^M$ of our model system is about 320 mV (see Fig.~\ref{fgr:dividing}b). Comparing to $\chi^0$ estimated from the $\gamma_T$  field-derivative (210 mv) one could argue that the similarity in value is sufficiently close that there could be relation between these quantities, or going further, that a good part of the electromechanical response coefficient $\chi^0$ is due to the spontaneous dipole potential. Note that this match is more difficult to make for the full potential (Fig.~\ref{fgr:dividing}c) because of the opposite sign. However, as discussed in the next section, this can not be the whole story. There must be additional contributions to $\chi^0$ not included in our simple membrane model.

Furthermore, the continuum model enables us to make an estimation of $\epsilon$ of the liquid water vapour interface using Eq.~\ref{eq:epsilon}. Given the  $l = 0.9$ nm used in our determination of $\gamma_T$ (Fig.~\ref{fgr:derivative}) we obtain a dielectric constant of $\epsilon \sim 3$, which is in accord with the low dielectric constant of water confined in the hydrophobic environment because of the dielectric dead-layer effect ~\cite{Zhang:2018tl}. However, because of the drastic approximations in the continuum model this value should be again treated with caution. For example the dielectric tensor in the interface is likely to be anisotropic with a different value in the normal and tangential direction\cite{Netz2012}.  

We now return to the system size dependence displayed in Fig.~\ref{fgr:derivative}. Recalling that the lateral dimensions $L_x= L_y$ of the MD cell remain fixed one would expect $1/L_z$ behaviour. This can be traced back to the box size dependence of the polarization interacting with the electric field it generates (the self field). There is in fact an explicit polarization self interaction in the expression Eq.~\ref{eq:uvdbM} of the MD constant Hamiltonian, namely the last term. This term scales as $1/L_z$ because the variance $\langle M^2 \rangle $ of the slab dipole converges to a constant (the dimensions of the water slab don't change). However the constant field extension is not explicitly included in the $\gamma_T$ of Fig.~\ref{fgr:derivative} which only shows the contribution of the Ewald Hamiltonian $H_{\mathrm{PBC}}$. Still, part of the polarization self interaction is already accounted for by the PBC Hamiltonian. This fraction will increase as the contribution of the $M^2/L_z$ term due to the extension decreases. Comparing to the results for the medium size box (green curve in Fig.~\ref{fgr:derivative}) we can assume that the curve for the large box (in red) are sufficiently close to convergence for the purpose of the analysis in the present paper.   

\section{Discussion and outlook}
\label{sec:disc}
The membrane model of section \ref{sec:dielbrane} is the most elementary of continuum models for the electromechanical response of the vapour-liquid interface. Its main merit is that it provides an explanation for the inverted parabolic variation of the lateral component $\gamma_T$ of the surface tension in response to a normally applied electric field. It achieves this by relating $\gamma_T$ to the Maxwell stress $\sigma^{\mathrm{M}}_T$ in the direction perpendicular to the applied field (parallel to the surface)\cite{Suo2008}.  Due to geometric coupling  $\sigma^{\mathrm{M}}_T$  is quadratic in the electric field with  negative curvature in agreement with the simulation results. 

The model also enabled us to interpret the observed shift of the maximum of the parabola away from zero electric field as a signature of spontaneous polarization. In fact in our simple treatment the finite zero field derivative of the lateral surface tension is directly equal to the surface potential generated by the spontaneous polarization (Eq.~\ref{eq:chi0}). Multiplication by a material dependent constitutive coefficient is not needed. This surprising and rather suspicious feature is a consequence of the generic character of geometric coupling and can be expected to disappear when additional electromechanical mechanisms are taken into account such as electrostriction (see below). Eq.~\ref{eq:chi0} has therefore not the status of a thermodynamic law similar to the Lippmann equation relating the potential derivative of the surface tension to the surface charge. Instead Eq.~\ref{eq:chi0} must regarded as a limiting constitutive relation.     

While capturing the leading mechanism for electromechanical coupling in deformable dielectric bodies, the model clearly has a number of severe limitations when extended to the liquid water vapour interface. First of all, the spontaneous polarization $P_0$ was introduced in Eq.~\ref{eq:EDP0} in a completely ad hoc fashion with the drastic approximation that $P_0$ is constant under a change of applied field (Eq.~\ref{eq:dP0dD}). As pointed out in the introduction the question of the origin of the spontaneous polarization in the thin liquid water vapour interface layer remains open. However, any such mechanism is likely to leave its imprint on the field derivative, which is a further reason why Eq.~\ref{eq:chi0} must be considered an approximation. The same question arises concerning the effect of electrostriction which we also left out. Similar to the Maxwell stress the contribution of electrostriction to the stress tensor is quadratic in the electric field (see again Eq.~\ref{eq:MHelmK}). According to the estimate of Ref.~\citenum{Mitra2013} the effect can be significant. Note, however that the sign of the curvature due to electrostriction is opposite to that of the geometric coupling assuming $\partial \epsilon / \partial \rho > 0 $, which is the normal behaviour for dielectric bulk material.

From a more fundamental perspective, the most serious approximation is undoubtedly that our simple membrane model treats the interface as a uniform slab  even though the root cause of surface tension is precisely the steep change in density\cite{Rowlinson1982}.  As touched upon earlier, VanderWaals theory accounts for this by adding a square gradient term to the local free energy density. Extending the free energy density $f_0(\rho)$ of Eq.~\ref{eq:fideal} with a similar square gradient term will provide a mechanism for the density to relax  under finite field ($\partial \rho /\partial D \ne 0$) which is likely to alter the field dependence of the surface tension and hence also the zero field derivative (Eq.~\ref{eq:chi0}). There exists already a literature on application of VanderWaals square gradient theory to polar fluid interfaces\cite{Onuki2004,Tsori2011,Mitra2013}. However, to the best of our knowledge, electromechanical coupling in these studies is entirely based on the density (concentration) dependence of the dielectric constant (electrostriction). Geometric coupling is a separate electromechanical mechanism\cite{Ogden2005,Suo2008}.  Without it the Maxwell stress effect is missed. In contrast, as already indicated in the introduction, geometric coupling is central in electroelastic strain gradient theory. For an example closely related to our system we refer to the work of Deng et al.\cite{Sharma2014} The formalism of Ref.~\citenum{Sharma2014} is again a Lagrangian theory using a material reference frame. While it remains to be seen whether this is necessary or convenient for the liquid vapour interface a similar deformation scheme with an appropriate liquid-like constitutive relation should be feasible.

The success of such an electroelastic description will depend of course  on the details of the constitutive model. In particular, while flexoelectricity is a tempting option worth investigating, it can not offer an universal explanation of spontaneous polarization at the liquid vapour interface of polar fluids. From simulation studies\cite{Eggebrecht1987,Grest2015,Grest2017}, statistical mechanics\cite{Eggebrecht1987,Thompson1987} and DFT calculations\cite{TelodaGama1991,Oxtoby1993,TelodaGama2002} it is known that the dipoles at the vapour liquid interface of a simple polar (Stockmayer) fluid show little alignment normal to the surface. Indeed, as already mentioned,  DFT investigations of systems based on point dipole and quadrupole interactions seem to imply that the role of quadrupole coupling is crucial\cite{Gray1992,Gray1994,Dietrich1992,Dietrich1993,Warschavsky2002,Warschavsky2003}. In fact ordering of quadrupoles at the interface may be the driving force stabilizing spontaneous polarization.  However, an atomistic mechanism such as this can be used to refine an electromechanical constitutive model leading to better understanding of the thermodynamic implications.     

The other fundamental issue we left unresolved is the question of the very definition of surface tension for a liquid-vapour interface  subject to a finite electric field. The main complication here is that Maxwell stress is an electrostatic field property and therefore persists in vacuum. The pressure in the vapour phase, even at vanishing density, can therefore reach values well beyond ambient pressure at zero field. Moreover the pressure tensor can be anisotropic also in uniform phases. This suggests a parallel with an interface between elastic solids with the possibility of surface tension and surface stress being no longer identical. This is the point of view taken by  Koski et al.~in their study of the liquid vapour interface of the Stockmayer fluid\cite{Grest2017}. It is therefore more than likely that the expression of Eq.~\ref{eq:gamplan} valid for non-polar fluids will need to be amended. To pursue this issue further technical development will also be necessary. Due to difficulties created by long-range electrostatic forces the method we used for the pressure tensor calculation was restricted to the parallel component of flat surfaces\cite{Sega2016}. Lifting this limitation is probably the number one priority for future continuation of this study.  

We end with a brief word on a possible link to the electrochemical surface potential $\chi^S$ of liquid water. The commonly quoted current value of this elusive quantity is 150 mV pointing inward from the vapour phase to the liquid\cite{Fawcett2008,Kristalik2008}. The consensus on this value was reached after reconciliation between ionic work functions as determined by electrochemical\cite{Fawcett2008,Kristalik2008} and molecular beam methods\cite{Tissandier1998}(See also Ref.~\citenum{Truhlar2006}). This value is qualitatively consistent with the surface dipole potential $\chi^M$ of ISM simulation. The numerical discrepancy of a factor two for SPC/E (150 vs 320 mv) while significant, is still sufficiently close to suggest a connection. This coincidence has led to a long discussion in the modelling community which continues till today\cite{Pratt1988,Pratt1992,Tildesley1997,Ichiye2015,Beck2013,Mundy2017,Hunenberger2018}.

With this paper we have added a third quantity $\chi^0$ in the comparison, the zero field derivative of the surface tension, which can be regarded as an electromechanical surface potential. The major advantage is that it is a response coefficient which is in principle accessible to experiment.  The (highly) preliminary estimate of 210 mV we obtained is between the electrochemical potential of experiment (150 mV) and the SPC/E dipole potential (320 meV). In our view, it is not clear whether any of these three surface potentials should apriori be equal to each other. One could conclude therefore that we have increased the confusion. On the positive side, what this work has contributed is that it opened up a new perspective on the water vapour interface potential as an electrocapillary effect with a still rather hypothetical derivative response relation to the surface tension

\section*{Conflicts of interest}
There are no conflicts to declare.

\section*{Acknowledgements}

CZ thanks M. Sega for helpful discussions about his virial version of GROMACS.  Steve Cox is acknowledged for discussion about the theory.



\balance



\providecommand*{\mcitethebibliography}{\thebibliography}
\csname @ifundefined\endcsname{endmcitethebibliography}
{\let\endmcitethebibliography\endthebibliography}{}

\bibliographystyle{rsc} 

\end{document}